\documentstyle[sprocl,epsfig]{article}

\def\Journal#1#2#3#4{{#1} {\bf #2}, #3 (#4)}

\def\be{\begin{equation}}
\def\ee{\end{equation}}
\def\bea{\begin{eqnarray}}
\def\eea{\end{eqnarray}}

\begin{document}

\title{The XMM-Newton Bright Serendipitous Source Sample}

\author{R. DELLA CECA}

\address{Osservatorio Astronomico di Brera, Via Brera 28, \\I-20121, 
Milano, Italy\\E-mail: rdc@brera.mi.astro.it} 

\author{ON THE BEHALF OF THE XMM-NEWTON SURVEY SCIENCE CENTRE (SSC) CONSORTIUM
\footnote{The SSC is 
an international collaboration involving 
a consortium of 
several institutions, appointed by  ESA, to exploit the XMM-Newton serendipitous 
detections for the benefit of the international scientific community
(see Watson et al., 2001 and http://xmmssc-www.star.le.ac.uk for a 
full description of the program).}
}


\maketitle

\vskip 0.5truecm

\abstracts{ We summarize here the current status of the 
XMM BSS: a large ($\sim 1000$ sources) 
sample of bright 
serendipitous XMM sources at high galactic latitude ($|b| > 20$ deg).
}

\section{Introduction}

Deep {\it Chandra} and {\it XMM--Newton} 
observations (Brandt et al. 2001; Rosati et al. 2002; Moretti et al., 2002;
Hasinger et al., 2001)
have recently resolved $> \sim 80\%$  of the 2--10 keV 
X-ray background (XRB) into discrete sources down to 
$f_{x}\sim 3\times$10$^{-16}$ erg cm$^{-2}$ s$^{-1}$.
The statistical analysis (stacked spectra and hardness ratios) 
performed on these samples provided an indication of the 
X--ray spectral properties of the sources making up most of the XRB.
The X--ray data are consistent with AGN 
being the dominant population of the XRB and, as 
inferred by the X--ray colors, a significant fraction 
of these sources have hard, presumably obscured, X--ray spectra, 
in agreement with the predictions of XRB synthesis models 
(see Madau et al., 1994; Comastri et al., 1995; Gilli et al. 2001).
However the majority of the sources found in these medium-deep fields 
are too faint to provide good X--ray spectral information. 
Moreover, the extremely faint magnitude  of a large part of their
optical counterparts makes the spectroscopic identifications
very difficult, or even impossible,
with the exisisting ground--based optical telescopes.
Thus, notwithstanding the remarkable results obtained by reaching very faint
X--ray fluxes, the broad--band physical  properties (e.g. the relationship 
between optical absorption and X-ray obscuration and the reason why AGN with 
similar X-ray properties have completely different optical 
appearance) are not yet completely understood.
A step forward toward the solutions of these issues has been recently
obtained by Mainieri et al., 2002 and Piconcelli  et al., 2002 using 
a small sample of serendipitous sources for which medium-good quality 
XMM-Newton and optical data are available. 
 
With the aim of complementing the results obtained by medium-deep X-ray
surveys, the XMM-Newton Survey Science Centre (SSC)
is building up the {\bf ``The XMM-Newton Bright Serendipitous Source Sample"} 
(XMM BSS, Della Ceca et al. 2002): a large ($\sim1000$ sources) 
sample of bright ($f_x\geq \sim 10^{-13}$ erg cm$^{-2}$ s$^{-1}$)  
serendipitous XMM sources at high galactic latitude ($|b| > 20^o$).
The well defined criteria (completeness, 
representativeness, etc..) of this sample will allow both a detailed study of sources 
of high individual interest and statistical population  studies.
In particular, the XMM BSS 
{\bf will be fundamental 
to complement other medium and deep XMM and {\it Chandra} survey 
programs (having fluxes 10 to 100 times fainter and covering a 
smaller area of the sky) and 
will provide a larger baseline for all evolutionary studies.}
Moreover, the high X--ray statistics which characterize most of the 
sources in the XMM BSS sample, combined with the  
relative  brightness of their optical counterparts, allow us to 
investigate in detail their physical properties.

\section*{The XMM BSS Sample}

The XMM BSS 
is part of the follow-up program being  conducted by the 
XMM-Newton Survey Science Center.
The XMM BSS is lead by the
{\it Osservatorio Astronomico di Brera} (Milan, Italy) and  consists of two 
flux-limited samples
having flux limits of $\sim 10^{-13}$ erg cm$^{-2}$ s$^{-1}$  
in the 0.5--4.5 keV (XMM BSS ``soft" sample) and in the 4.5-7.5 keV 
(XMM BSS ``hard" sample) energy band. 

As of today, 195 suitable XMM-Newton  fields have been analyzed and a  first 
sample of 331 sources selected: 321 sources belongs to the 
``soft" sample and 64 sources to the ``hard" sample with 
54 sources in common.
The optical counterpart of the  majority (85-90\%) of these X-ray sources 
has an optical magnitude above the POSS II limit (R $\sim 21^{mag}$), thus 
allowing spectroscopic identification on a 4 meter telescope.
It is worth noting that, given the accuracy  
of the X-ray positions (2-5 arcsec 
at the 90\% confidence level) and the magnitude of the expected optical 
counterpart, only one object needs  to be observed to obtain the optical 
identification. For this reason the complete spectroscopic 
identification of the two samples is feasible with a reasonable 
number of telescope nights.
Up to now 177 sources have been spectroscopically 
identified (either from the literature or from our own
observations at ESO, TNG and Calar Alto telescopes) 
leading to a 54\% and 73\% identification
rate for the ``soft'' and ``hard'' samples respectively. 
The optical breakdown of the XMM BSS sources identified so far 
is reported in Table 1.

\begin{center}
Table 1: The current optical breakdown of the XMM BSS Sample\\
\begin{tabular}{lrr}
\hline
\hline
 &  ``Soft" Sample                      & ``Hard" Sample  \\			  
 & $S_{0.5-4.5 keV} \geq 7\times 10^{-14}$ cgs  &  $S_{4.5-7.5 keV} \geq 1\times 10^{-13}$ cgs\\
\hline
Objects$^{1}$ & 321  & 64  \\
\\
Identified: & 172  & 47  \\
AGN-1 & 113  & 26  \\
AGN-2 & 19   & 15   \\
Clusters of Galaxies & 3   & 1   \\
Galaxies & 6   & 3   \\
BL Lacs & 2   & 0   \\
Stars & 29  & 2   \\
\hline
\hline 
\end{tabular}
\end{center}
$^{1}$ Note that 54 sources are in common between the ``soft" and ``hard" sample.

\section*{The hardness ratio distribution}

A ``complete" spectral analysis for all the sources in the 
XMM BSS is in progress; in the meantime 
a ``snapshot" of the X-ray spectral properties of the identified 
sources 
obtained using the ``hardness ratio" method (equivalent to 
the ``color-color" analysis largely used at optical wavelengths)
is shown in figure 1.   
A fairly sharp separation between Galactic and extragalactic
sources is visible in figure 1 (left panel):
25 out of 29 stars have HR2$<$--0.7.
Moreover it is worth noting that both in the hard and in the soft samples 
Broad Line AGNs lie in the range 
-0.8$<$HR2$<$--0.3 (except for a few cases). On the contrary 
Narrow Line AGNs are distributed over a larger area in the 
``hardness ratio" plot with the trend to have a larger HR2 value 
for the Narrow Line AGNs belonging to the XMM BSS ``hard" sample.
Besides the theoretical implications of this 
segregation, the sensitivity of the HR2 value to the 
optical spectral type of the X-ray sources can offer 
a powerful tool to increase the efficiency for the selection
of rare and interesting classes of objects (e.g. the absorbed 
AGNs).
Preliminary results for a first sample of ``XMM BSS optically dull"
galaxies are discussed in Severgnini et al., (this conference). 

\begin{figure}[t!]
\vskip -3truecm
{\centerline {\epsfig{figure=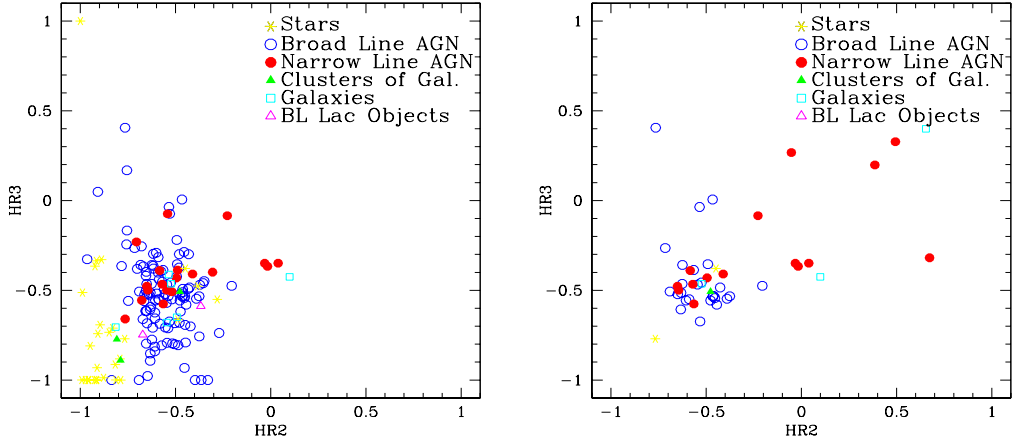, width=15cm, height=10cm, angle=-90}}}
\vskip -2truecm
\caption{HR2 vs. HR3 for the identified 
objects belonging to the XMM BSS ``soft" sample (left 
panel) and to the XMM BSS ``hard" sample (right panel).
HR2 and HR3 for each source have been computed using the
source count rate in the (0.5$-$2\, {\rm keV}),
(2$-$4.5\, {\rm keV}) and (4.5$-$7.5\, {\rm keV}) energy band according
to:
$HR2=[C(2-4.5)-C(0.5-2)]/[C(2-4.5)+C(0.5-2)]$
and
$HR3=[C(4.5-7.5)-C(2-4.5)]/[C(4.5-7.5)+C(2-4.5)]$.
We have used different colors to mark the identified objects and,
for clarity, we have not reported error bars.
} 
\end{figure} 

\section*{Acknowledgments}
This work has received financial support from ASI(I/R/037/01) and from 
MURST (Cofin 00-02-004). We thank the ESO, TNG and Calar Alto 
telescope allocation committee for supporting this project.

\end{document}